\magnification=1200 \vsize=25truecm \hsize=16truecm \baselineskip=0.6truecm
\parindent=1truecm \nopagenumbers \font\scap=cmcsc10 \hfuzz=0.8truecm

\null \bigskip  \centerline{\bf CELLULAR AUTOMATA AND ULTRA-DISCRETE
PAINLEV\'E EQUATIONS}
\vskip 2truecm

\centerline{\scap B. Grammaticos}
\centerline{\sl LPN, Universit\'e Paris VII}
\centerline{\sl Tour 24-14, 5${}^{\grave eme}$\'etage}
\centerline{\sl 75251 Paris, France}
\bigskip
\centerline{\scap Y. Ohta}
\centerline{\sl Department of Applied Mathematics}
\centerline{\sl Faculty of Engineering, Hiroshima University}
\centerline{\sl 1-4-1 Kagamiyama, Higashi-Hiroshima 739, Japan}
\bigskip
\centerline{\scap A. Ramani}
\centerline{\sl CPT, Ecole Polytechnique}
\centerline{\sl CNRS, UPR 14}
\centerline{\sl 91128 Palaiseau, France}
\bigskip
\centerline{\scap D. Takahashi}
\centerline{\sl Department of Applied Mathematics and Informatics}
\centerline{\sl Ryukoku University}
\centerline{\sl Seta, Ohtsu 520-21, Japan}
\bigskip
\centerline{\scap K.M. Tamizhmani}
\centerline{\sl Departement of Mathematics}
\centerline{\sl Pondicherry University}
\centerline{\sl Kalapet, Pondicherry, 605104 India}
\bigskip

\vskip 2truecm \noindent Abstract \smallskip
\noindent  Starting from integrable cellular automata we present a novel
form of Painlev\'e
equations. These equations are discrete in both the independent variable
and the dependent one.
We show that they capture the essence of the behavior of the Painlev\'e
equations organizing
themselves into a coalescence cascade and possessing special solutions. A
necessary condition for
the integrability of cellular automata is also presented.
\vfill\eject

\footline={\hfill\folio} \pageno=2

\medskip
\noindent A novel extension of the Painlev\'e equations, well-known for
their numerous
applications in mathematics and physics, will be presented in this paper.
Recent progress in the
domain of integrable discrete systems has led to the derivation of discrete
analogs of the
Painlev\'e equations [1]. They were first identified in a 2D model of
quantum gravity where they
appeared as an integrable recursion relation for the calculation of the
partition function [2]. With
hindsight, their first occurence can be traced back to the work of Jimbo
and Miwa on the spin-spin
correlation function of a 2D Ising lattice [3]. To date, the full list of
discrete Painlev\'e
equations has been established [4] and their properties (in perfect
parallel with those of their
continuous counterparts) are being actively investigated. Moreover the
study of discrete integrable
systems has made clear the fact that they, rather than the continuous
ones, are the fundamental entities. In fact, the latter are
contained, through appropriate limits, in the discrete ones and thus we can
establish the
hierarchical diagram:

Discrete systems $\to$ Semi-continuous systems $\to$ Continuous systems

\noindent A recent discovery has made possible to extend this structure in
another direction. In
[5] one of the present authors has proposed a systematic way for the
introduction of integrable
ultra-discrete systems. In previous works on discrete systems, while the
independent variables
were discrete, the {\sl dependent} variables were assumed to vary
continuously. The ultra-discrete
limit provides a systematic way to discretize the dependent variable. One
can, starting from
a given evolution equation, obtain the cellular automaton (CA) equivalent.
The aim of this paper is
two-fold. First, introduce the ultra-discrete analogs of the Painlev\'e
equations and investigate
their properties and, second, provide integrablility conditions for
cellular-automaton like
equations.

Cellular automata have been the object of an impressive number of studies
and their behavior is
known to be of the utmost richness. The integrability of such systems has
not been thoroughly
studied, since it represents considerable difficulties. An occurence of an
integrable automaton
has been noted in [6] by Pomeau who obtained explicitly its constant of
motion. Cellular automata
representing evolution equations have been studied from the point of view
of the existence of
localized, soliton-like, solutions. The notion of soliton for CA's was
first introduced by Park et
al. in [7]. Further examples of such CA's with soliton-like structures were
given by the Clarkson
group [8]. Integrable CA were introduced by Bruschi and collaborators [9]
who derived Lax pairs for
their cellular-automaton equations. Bobenko et al. [10] have proposed an
interesting approach to
integrable CA's by considering them as the restriction of an integrable
discrete equation over a
finite field.
 However in many cases the
relation to the well-known integrable evolution equations was based on
circumstancial evidence
rather than a systematic derivation. The situation has changed recently due
to the introduction of
a method [5] that allows one to convert a given {\sl discrete} evolution
equation to one where the
dependent variable also takes discrete values.
The starting point was the CA proposed by one of the authors in
collaboration with Satsuma [11].
This simple model (essentially a ``box and ball'' system) was shown later
to be the ultra-discrete
limit of the KdV equation [5]. This was obtained through a limiting procedure
(the ultra-discrete limit) which allows one to derive a cellular automaton
equation starting from
the appropriate form of a discrete evolution equation.

As an illustration of the method and a natural introduction to
ultra-discrete Painlev\'e equations,
let us consider the following discrete Toda system:
$$u_n^{t+1}-2u_n^t+u_n^{t-1}=
\log (1+\delta^2(e^{u_{n+1}^t}-1))-2\log (1+\delta^2(e^{u_{n}^t}-1))+\log
(1+\delta^2(e^{u_{n-1}^t}-1))\eqno(1)$$ which is the integrable
discretization of the continuous
Toda system:
$${d^2r_n\over dt^2}=e^{r_{n+1}}-2e^{r_{n}}+e^{r_{n-1}}\eqno(2)$$
We introduce $w$ through $\delta=e^{-L/2\epsilon}$, $w_n^t=\epsilon
u_n^t-L$ and take the limit
$\epsilon\to 0$. We use the well-known result $\lim_{\epsilon\to 0}\epsilon\log
(1+e^{x/\epsilon})={\rm max} (0,x)=(x+|x|)/2\equiv (x)_+$. The function
$(x)_+$ is also known under
the name of truncated power function and is equal to 0 for $x\leq 0$ and
$x$ for $x>0$. Thus the
ultra discrete limit of (1) becomes simply [12]:
$$w_n^{t+1}-2w_n^t+w_n^{t-1}=
(w_{n+1}^t)_+-2(w_{n}^t)_++(w_{n-1}^t)_+\eqno(3)$$
Since the truncated power function of any integer is a non-negative
integer, it is clear that if
$w$ has initially integer values the values will remain integer at all
subsequent time steps. Thus
equation (3) is indeed a cellular automaton equation.

Let us now restrict ourselves to a simple periodic case with period two
i.e. $r_{n+2}=r_n$ and
similarly $w_{n+2}=w_n$. Calling $r_0=x$ and $r_1=y$ we have from (2) the
equation $\ddot
x=2e^{y}-2e^{x}$ and $\ddot y=2e^{x}-2e^{y}$ resulting to $\ddot x+\ddot
y=0$. Thus $x+y=\mu
t+\nu$ and we obtain after some elementary manipulations:
$$\ddot x=ae^{\mu t}e^{-x}-2e^{x}\eqno(4)$$
Equation (4) is a special form of the Painlev\'e P$_{\rm III}$ equation.
Indeed, putting
$v=e^{x-\mu t/2}$, we find:
$$\ddot v={\dot v^2\over v}+e^{\mu t/2} (a-2v^2)\eqno(5)$$
The same periodic reduction can be performed on the ultra-discrete Toda
equation (3). We
introduce $w_0^t=X^t$, $w_1^t=Y^t$ and have, in perfect analogy to the
continuous case,
$X^{t+1}-2X^t+X^{t-1}=2(Y^t)_+-2(X^t)_+$ and
$Y^{t+1}-2Y^t+Y^{t-1}=2(X^t)_+-2(Y^t)_+$. Again, $\Delta _t^2(X^t+Y^t)=0$
and we can take
$X^t+Y^t=mt+p$ (where $m,t,p$ take integer values). We find thus that $X$
obeys the ultra-discrete
equation:
$$X^{t+1}-2X^t+X^{t-1}=2(mt+p-X^t)_+-2(X^t)_+\eqno(6)$$
This is the ultra-discrete analog of the special form of the Painlev\'e
P$_{\rm III}$ equation
(5). Figure 1 gives a comparison of the evolution under (4) and (6). We
remark that the dynamics
are very similar: the two particles converge towards each other, rebound
once or twice, get
captured and go on oscillating around some equilibrium point. Thus,
starting from a well-known
physical problem we have introduced the corresponding cellular automaton
equation and, restricting
it to the simplest periodic lattice, we obtained the ultra-discrete form of
a Painlev\'e equation.

In order to construct the ultra-discrete analogs of the Painlev\'e
equations (u-P) we must start
with the discrete form that allows the ultra-discrete limit to be taken.
The general procedure is to
start with an equation for $x$, introduce $X$ through $x=e^{X/\epsilon}$
and then take
appropriately the limit $\epsilon\to 0$. Clearly the substitution
$x=e^{X/\epsilon}$ requires $x$
to be positive. This is a stringent requirement that limits the exploitable
form of the d-P's to
multiplicative ones. Fortunately many such forms are known for the discrete
Painlev\'e transcendents
[13]:

\noindent d-P$_{\rm I}$
$$x_{n+1}x_{n-1}={\alpha\lambda^n\over x_n} +{1\over x_n^2}\eqno(7a)$$
$$x_{n+1}x_{n-1}=\alpha\lambda^n +{1\over x_n}\eqno(7b)$$
$$x_{n+1}x_{n-1}=\alpha\lambda^nx_n+1\eqno(7c)$$
d-P$_{\rm II}$
$$x_{n+1}x_{n-1}={\lambda^n(x_n+\alpha\lambda^n)\over x_n(1+x_n)}\eqno(7d)$$
$$x_{n+1}x_{n-1}={x_n+\alpha\lambda^n\over 1+\beta x_n\lambda^n}\eqno(7e)$$
d-P$_{\rm III}$
$$x_{n+1}x_{n-1}={(x_n+\alpha\lambda^n)(x_n+\beta\lambda^n)\over
(1+\gamma x_n\lambda^n)(1+\delta x_n\lambda^n)}\eqno(7f)$$
We remark that in some cases, more than one form
exists for a given d-P.  The derivation of the equivalent ultra-discrete
forms is elementary: we
take the logarithm of both sides of the equation and whenever a sum appears
we apply the limit
leading to the truncated power function. We find thus:

\noindent u-P$_{\rm I}$
$$X_{n+1}+X_{n-1}+2X_n=(X_n+n+a)_+\eqno(8a)$$
$$X_{n+1}+X_{n-1}+X_n=(X_n+n+a)_+\eqno(8b)$$
$$X_{n+1}+X_{n-1}=(X_n+n+a)_+\eqno(8c)$$
u-P$_{\rm II}$
$$X_{n+1}+X_{n-1}=n+(n+a-X_n)_+-(X_n)_+\eqno(8d)$$
$$X_{n+1}+X_{n-1}-X_n=(n+a-X_n)_+-(X_n+n+b)_+\eqno(8e)$$
u-P$_{\rm III}$
$$X_{n+1}+X_{n-1}-2X_n=(n+a-X_n)_++(n+b-X_n)_+-(X_n+c+n)_+-(X_n+d+n)_+\eqno(
8f)$$
These equations describe celular automata provided we restrict the values
of the parameters as
well as the initial values of the dependent variable to integers. Figure 2
shows a comparison of the discrete P$_{\rm I}$
(7c) with the ultra-discrete P$_{\rm I}$ (8c). It is remarkable that the
behavior of the two
equations is very similar (provided that we plot the logarithm of the
variable of the discrete
equation, as expected).
The forms (8) are not canonical in the sense that they can be simplified by
a translation of $X$
and a linear transformation in $n$. We will not enter into these details
but merely list the
canonical forms obtained:

\noindent canonical u-P$_{\rm I}$
$$X_{n+1}+X_{n-1}+\sigma X_n=(X_n+n)_+\quad{\rm with\ \sigma=0,1,2}\eqno(9a)$$
canonical u-P$_{\rm II}$
$$X_{n+1}+X_{n-1}-\sigma X_n=a+(n-X_n)_+-(n+X_n)_+\quad{\rm with\
\sigma=0,1}\eqno(9b)$$
canonical u-P$_{\rm III}$
$$X_{n+1}+X_{n-1}-2X_n=(n+a-X_n)_++(n-a-X_n)_+-(X_n+b+n)_+-(X_n-b+n)_+\eqno(
9c)$$
Note that equation (6), for $m=2, p=0$ is the subcase $a=b=0$ of (9c) after
the change of variable
$X^t=X_n+n$.

  At this point two questions appear unavoidable. First, is it justified to
call these
equations ultra-discrete {\sl Painlev\'e} equations? What do they have in
common with the
familiar Painlev\'e equations? One first remark is that the u-P's form a
coalescence cascade just
like their continuous and discrete counterparts [13]. Indeed, starting from
u-P$_{\rm II}$ (8d)
(respectively (8e)) we can take
$b\to -\infty$ in which case the nonlinear term containing it is always
zero. We transform the
resulting equation using the identity $(x)_+=x+(-x)_+$, then translate $X$
and obtain equation
u-P$_{\rm I}$ in the form (8b), (resp. (8c)).
We can also recover (8a) from (8d) (and (8b) also from (8c)).
Similarly starting from u-P$_{\rm III}$
(8f) we can obtain u-P$_{\rm II}$ (8d) by taking $b\to +\infty$ and $c\to
+\infty$ such that
$b-c$ is finite. Next we translate $X$ and through a linear transformation
of $n$ we find
u-P$_{\rm II}$ (8d). This is not the only property the u-P's share with the
continuous
and discrete Painlev\'e equations as we shall see below. The second
question is whether it is
possible to guess the forms of the u-P's. In other words, what is the
(integrability)
criterion that would single them out among all possible equations?

In the case of discrete systems the criterion for integrability (equivalent
to the Painlev\'e
property) is the property known as singularity confinement [14]. For
cellular automata no
singularity can exist and thus this criterion is inoperable. The situation
is analogous to
polynomial mappings where no singularity can appear. There, the criterion
for integrability is
based on arguments of growth of the degree of the iterate (or the similar
notion of complexity
introduced by Arnold [15]). Veselov [16] has shown that among mappings of
the form
$x_{n+1}-2x_n+x_{n-1}=P(x_n)$ with polynomial $P(x)$, only the linear one
has non-exponential
growth of the degree of the polynomial that results from the iteration of
the initial conditions.
Let us apply such a low-growth criterion to a family of ultra-discrete
P$_{\rm I}$ equations of the
form:
$$X_{n+1}+\sigma X_n+X_{n-1}=(X_n+\phi(n))_+\eqno(10)$$
The three u-P$_{\rm I}$ obtained from (7) correspond to $\sigma =0,1,2$.
What is the condition for
$X$ not to grow exponentially towards $\pm\infty$? We ask simply that the
polynomials $r^2+\sigma
r+1$ and $r^2+(\sigma -1)r+1$ have complex roots (otherwise exponential
growth ensues). The only
{\sl integer} values of $\sigma$ satisfying this criterion are $\sigma
=-1,0,1,2$. We remark that
the three values mentioned above are exactly retrieved plus the value
$\sigma =-1$. A close
inspection of this mapping shows that it is also integrable: it is just a
form of an
ultra-discrete  P$_{\rm III}$, obtained from the discrete system
$x_{n+1}x_{n-1}=x_n(x_n+\lambda^n)$ which leads to (10) with $\phi(n)=0$.

We have applied the low-growth criterion to other cases like u-P$_{\rm II}$
and u-P$_{\rm III}$ and
in every case the results of the growth analysis correspond to the already
obtained integrable
cases. However low-growth is not a sufficiently powerful integrability
criterion. In particular the
inhomogeneous terms ($\phi$ in equation (10)) cannot be fixed by this
argument. Any
$\phi (n)$ would satisfy this requirement. So another criterion must
complement this first one.

In the case of (continuous) evolution equations two integrability criteria
are often used in
conjunction: the Painlev\'e property and the existence of multisoliton
solutions. In the case of
Painlev\'e equations the latter are the special solutions that exist for
particular values of the
parameters [17] (except for P$_{\rm I}$ which is parameter-free). A
particular class of these
solutions (existing also in the discrete case) are the rational ones. We
shall investigate this
property in the case of u-P equations. This will strengthen the argument
that (9) are indeed
Painlev\'e equations and will allow us to fix completely their form. For
d-P$_{\rm II}$ the
simplest rational solution is a constant. Thus a constant solution should
exist for u-P$_{\rm II}$
and indeed for (9b) with $\sigma=0$ we find that $X=0$ is       a solution
for $a=0$. However this
solution exists if we replace $n$ in (9b) by any function of $n$. The next
solution
for  u-P$_{\rm II}$ is a step-function one. Indeed when $n$ is large
constant positive solution
exists with $X$ equal to
$a/4$ while a constant solution with $X=a/2$ exists when $n$ is large negative.
Thus, for instance for $a=4$ a solution for $n<<0$ is $X=2$, while for
$n>>0$ a solution is $X=1$.
The remarkable fact is that that these constant ``half" solutions do really
join to form a solution
of (9b)  with a unique jump at $n=-1$.
It is straightforward to check that this will not be the case in general if
the non-autonomous part
is not linear in $n$. The general solution of this type becomes now clear. For
$a=4m$ we have a solution with $m$ jumps from the value $X=2m$ to $X=m$.
The first jump occurs at
$n_0=1-2|m|$ and we have
successive jumps of $-|m|/m$  at $n=n_0+3k$, $k=0,1,2,\dots,|m|-1$.
Analogous results can be obtained for the other u-P$_{\rm II}$
corresponding to $\sigma =0$. Thus
u-P$_{\rm II}$ has a rich structure of particular solutions.

On the basis of these results we can conclude that the integrability
criterion for automata-like
Painlev\'e equations  appears to be based on low-growth arguments together
with the existence of a
rich class of explicit, globally described, solutions. The need for such a
two-step process is the
price we have to pay in the CA case because of the absence of
singularities. Can these ideas be
applied to other CA evolution equations? Let us present the example of
ultra-discrete Burgers
equation:
$$X^{t+1}_n=X^{t}_n+(X^{t}_{n+1})_+-(X^{t}_n)_+\eqno(11)$$
Using the identity $(X)_+=X+(-X)_+$ we can rewrite (11) as:
$$X^{t+1}_n=a(X^{t}_{n+1})_+-b(-X^{t}_n)_+\eqno(12)$$
where, from (11), $a=b=1$. How can we obtain these values of the parameters
based on a low-growth
argument? If either $|a|>1$ or $|b|>1$ we can find a special direction where
either $X^{t+k}_n$ or $X^{t+k+1}_{n-k}$ grows exponentially. Thus the only
{\sl integer} values that
$a$ and
$b$ can take are $\pm 1$ and 0. The value 0 leads to uninteresting, one
dimensional evolution and
the case $a=b=-1$ is equivalent to the case $a=b=1$. Thus the only other
case that cannot be
excluded on low-growth arguments is the case $a=1$ and $b=-1$. Although the
discrete equation from
which we could have obtained it is not integrable the behavior of the
automaton-like equation is
regular and there is as yet no  indication as to its nonintegrability.

It is clear from the results presented in [5] and the ones above that we
can produce
cellular-automaton-like equations starting from integranble nonlinear
evolution equations. The
present work has focused on the ultra-discrete forms of Painlev\'e
equations of which the first
three were given. The forms of the remaining u-P equations  will also be
investigated: probably
they will require a two-component description. The properties of their
solutions as well as the
relations between the various u-P's must also be studied. These open
questions will be addressed in
future works. What is important at this stage is that we have shown that
this new domain of
integrable systems is particularly rich. While the discrete systems are the
fundamental entities
and contain all the structure, the cellular automata are their bare-bones
version capturing the
essence of the dynamics. This explains the interest that these
ultra-discrete systems present both
from the mathematical and the physical points of view.

\vfill\eject
{\scap Acknowledgements}.
\smallskip\noindent
The authors wish to acknowledge the financial support of the CEFIPRA (under
contract 1201-1)
that had made the present collaboration possible.
Y. Ohta acknowledges the financial support of the Japan Ministry of
Education through the Foreign Study Program.
\noindent
\smallskip {\scap References}.
\smallskip
\item{[1]} A. Ramani, B. Grammaticos and J. Hietarinta, Phys. Rev. Lett. 67
(1991) 1829.
\item{[2]}      E. Br\'ezin and V.A. Kazakov, Phys. Lett. 236B (1990) 144.
\item{[3]}      M. Jimbo and T. Miwa, Proc. Japan Acad. 56 (1980) 405.
\item{[4]} B. Grammaticos and A. Ramani, {\sl Retracing the
Painlev\'e-Gambier classification for
discrete systems}, to appear in Meth. Appl. Anal (1996).
\item{[5]} T. Tokihiro, D. Takahashi, J. Matsukidaira and J. Satsuma, {\sl
From soliton equations to
integrable Cellular Automata through a limiting procedure}, preprint (1996).
\item{[6]} Y. Pomeau, J. Phys. A 17 (1984) L415.
\item{[7]} K. Park, K. Steiglitz and W.P. Thurston, Physica 19D (1986) 423.
\item{[8]} T.S. Papatheodorou, M.J. Ablowitz and Y.G. Saridakis, Stud.
Appl. Math. 79 (1988) 173.

T.S Papatheodorou and A.S. Fokas, Stud. Appl. Math. 80 (1989) 165.
\item{[9]} M. Bruschi, P.M. Santini and O. Ragnisco, Phys. Lett. A 169
(1992) 151.
\item{[10]} A. Bobenko, M. Bordemann, C. Gunn and U. Pinkall, Commun. Math.
Phys. 158 (1993) 127.
\item{[11]} D. Takahashi and J. Satsuma, J. Phys. Soc. Jpn. 59 (1990) 3514.
\item{[12]} D. Takahashi and J. Matsukidaira, Phys. Lett. A 209 (1995) 184.
\item{[13]} A. Ramani and B. Grammaticos, {\sl Discrete Painlev\'e
equations: coalescences, limits
and degeneracies}, to appear in Physica A (1996).
\item{[14]} B. Grammaticos, A. Ramani and V. Papageorgiou, Phys. Rev. Lett.
67 (1991) 1825.
\item{[15]} V.I. Arnold, Bol. Soc. Bras. Mat. 21 (1990) 1.
\item{[16]} A.P. Veselov, Comm. Math. Phys. 145 (1992) 181.
\item{[17]} V.I. Gromak and N.A. Lukashevich, {\sl The analytic solutions
of the Painlev\'e
equations}, (Universitetskoye Publishers, Minsk 1990), in Russian.

\bigskip
\noindent{\scap Figure captions}.
\smallskip
\item{Figure 1.} Distance between the two particles as a function of time
in the case (a) of the
continuous Toda potential and (b) its ultra-discrete analog.
\item{Figure 2.} Solution of the discrete Painlev\'e I equation (a)  and of
its ultra-discrete
analog (b).

\end